\documentclass{appolb}
\usepackage{epsfig}

\begin{document}
\title{Toward the description of fluid dynamical fluctuations in heavy-ion collisions%
\thanks{Presented by M.~Nahrgang at Critical Point and Onset of Deconfinement (CPOD) 2016}%
}

\author{Marlene Nahrgang$^{1,3}$, 
Marcus Bluhm$^{2,4}$, Thomas Sch\"{a}fer$^{4}$, Steffen Bass$^{3}$
\address{$^1$ SUBATECH UMR 6457 (IMT Atlantique, Universit\'e de Nantes,
IN2P3/CNRS), 4 rue Alfred Kastler, 44307 Nantes, France\\
$^2$ Institute of Theoretical Physics, University of Wroclaw, 50204 Wroclaw, Poland\\
$^3$ Department of Physics, Duke University, Durham, NC 27708-0305, USA\\
$^4$ Physics Department, North Carolina State University, Raleigh, NC 27695, USA}
}

\maketitle

\begin{abstract}
In this talk we present results obtained when fluid dynamical fluctuations are included in relativistic 
$3+1$ dimensional viscous fluid dynamics. We discuss effects of the interactions of fluctuations due to 
nonlinearities and the cutoff dependence.
\end{abstract}

  
\section{Introduction}

Fluid dynamics has become the prefered tool to describe the bulk physics of heavy-ion collisions. After a pre-equilibrium stage the initial densities of energy, momentum and conserved charges, which make up the first instances of the quark-gluon plasma (QGP) are propagated according to relativistic fluid dynamics until particlization to a dilute hadronic gas. Including the QCD equation of state obtained from lattice QCD calculations as well as shear and bulk viscosities various models have been able to describe the measured particle spectra and collective flow harmonics.

Despite this success there are a couple of issues with this fluid dynamical approach that need refinement before we can claim the precision extraction of fundamental QGP properties. In this talk we address recent progress that has been made on one of these issues which is of particular importance for studies of the QCD phase transition and for smaller systems: fluid dynamical fluctuations.

Conventional fluid dynamics propagates only thermal averages of energy, momentum and charge densities. We know, however, that even in thermal equilibrium fluctuations are present which depend on the number of particles or the volume. The inclusion of dissipative processes into the fluid dynamical equations via the shear-stress tensor and the bulk pressure, force us -- according to the fluctuation dissipation theorem -- to take fluctuations into account as well. At the QCD phase transition, in particular near the conjectured critical point, fluctuation phenomena are expected to be of crucial significance for experimental signals. These reasons urge us to develop models for the propagation of fluctuations which can be applied to describe the dynamics and nonlinearities of critical phenomena. 

First models \cite{Nahrgang:2011mg,Nahrgang:2011mv,Nahrgang:2011vn,Herold:2013bi,Herold:2016uvv} to couple the propagation of fluctuations with the fluid dynamical evolution in heavy-ion collisions have been based on effective models of QCD, like the quark-meson model. The chiral condensate, as the order parameter for chiral symmetry breaking, is propagated explicitly in these models and fluctuations couple to the fluid dynamical evolution via a stochastic source term. It has been found that critical slowing down weakens fluctuation signals in dynamical models compared to calculations in static, grandcanonical systems. After particlization critical fluctuations are imprinted in the net-proton number.

This talk discusses the fluid dynamical propagation of intrinsic fluctuations, which are derived in a model-independent approach.

The effect of fluctuations on fluid dynamics has before been studied either in linearized Bjorken expansions \cite{Kapusta:2011gt,Kapusta:2012zb,Akamatsu:2016llw} or in linearized numerical implementations \cite{Young:2014pka}. First attempts at a fully numerical implementation \cite{Murase:2016rhl} show an impact on flow harmonics but lack an explicit discussion of crucial aspects like the cutoff dependence of nonlinear interactions.

\section{Fluid dynamical fluctuations}
Starting from the linearized fluid dynamical equations for relativistic $3+1$ dimensional systems, the correlators of the noise terms can be obtained in linear response theory. For a noise field tensor $\Xi^{\mu\nu}$ feeding into the full fluid dynamical equations
\begin{equation}
 \partial_\mu T^{\mu\nu}=\partial_\mu\left(T^{\mu\nu}_{\rm eq} +  \Delta T^{\mu\nu}_{\rm visc} + \Xi^{\mu\nu}\right) = 0 \, ,
\label{eq:fluideq}
\end{equation}
one obtains in the Gaussian white noise approximation
\begin{eqnarray}
\label{eq:noiseav}
 \langle\Xi^{\mu\nu}\rangle &=& 0\, ,\\
\label{eq:covariance}
\langle\Xi^{\mu\nu}(x)\,\Xi^{\alpha\beta}(x')\rangle &=& 2T\left(\eta(\Delta^{\mu\alpha}\Delta^{\nu\beta}+\Delta^{\mu\beta}\Delta^{\nu\alpha})\nonumber\right.\\
                                &&\left.+(\zeta-2\eta/3)\Delta^{\mu\nu}\Delta^{\alpha\beta}\right)\delta^4(x-x')\, .
\end{eqnarray}
In order to overcome the acausality in the relativistic Navier-Stokes equations, the shear-stress tensor and the bulk pressure contained in $\Delta T_{\rm visc}^{\mu\nu}$ are themselves dynamical quantities. They evolve via the Israel-Stewart relaxation equations, which introduces relaxation times. In our approach we have chosen to evolve the noise fields analoguously to the viscous corrections via
\begin{equation}
 u^{\gamma}\partial_\gamma\Xi^{\langle\mu\nu\rangle}=-\frac{\Xi^{\mu\nu}-\xi_{\rm Gauss}^{\mu\nu}}{\tau_\pi}\, ,
\end{equation}
where $\xi_{\rm Gauss}^{\mu\nu}$ has the same covariances as the original noise tensor given in (\ref{eq:covariance}). This introduces time correlations in the noise fields $\Xi^{\mu\nu}$. In the example calculations shown in this work, we set $\zeta = 0$ and choose the noise relaxation time to be the same as the shear relaxation time $\tau_\pi$.

\begin{figure}
\centering
 \includegraphics[width=0.8\textwidth]{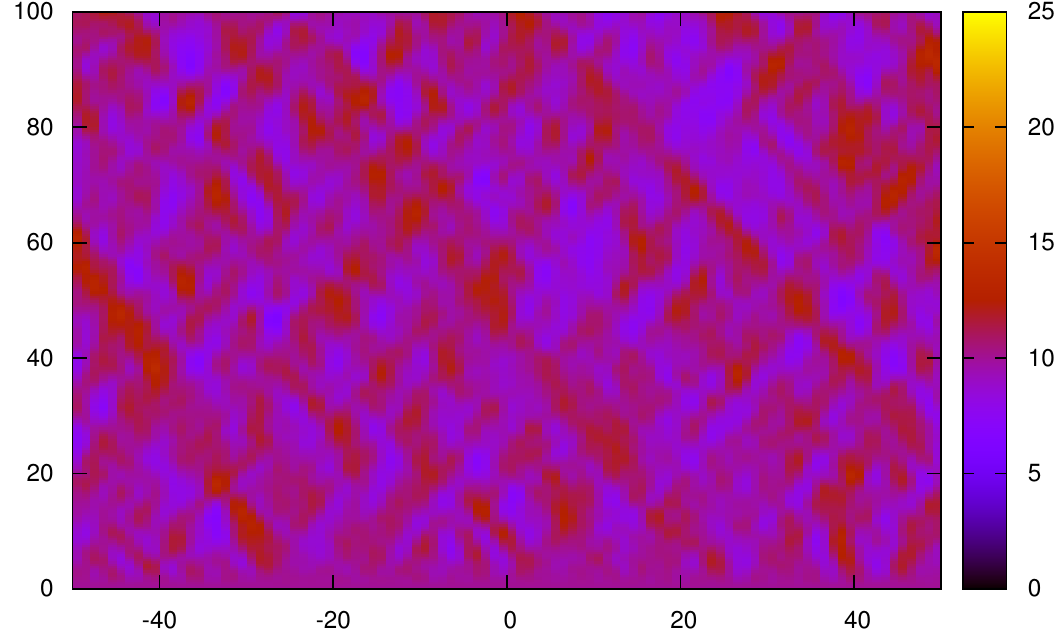}
\caption{Time evolution from $0-100$~fm/c of the energy density in GeV/fm$^3$ in one direction of the fluid confined in a box.}
\label{fig:edenscolormap}
\end{figure}
In the following we present calculations, where we have included fluctuations in the $3+1$d relativistic viscous fluid dynamics code vHLLE \cite{Karpenko:2013wva}.
Fig. \ref{fig:edenscolormap} shows the time evolution of the energy density $e$ along one direction of the fluid confined in a box with periodic boundary conditions. The noise field creates local fluctuations around the average energy and momentum density in the box, which are then transported according to the fluid dynamical equations. The lattice spacing used here is $\Delta x =1$~fm. We discuss the dependence on the lattice spacing below.

\begin{figure}
 \includegraphics[width=0.49\textwidth]{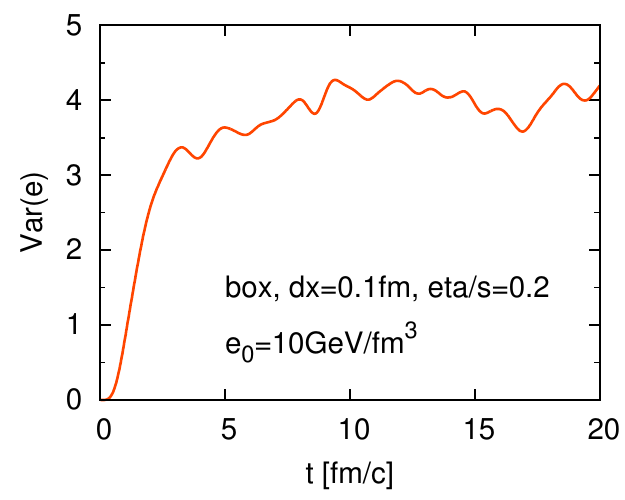}
 \includegraphics[width=0.49\textwidth]{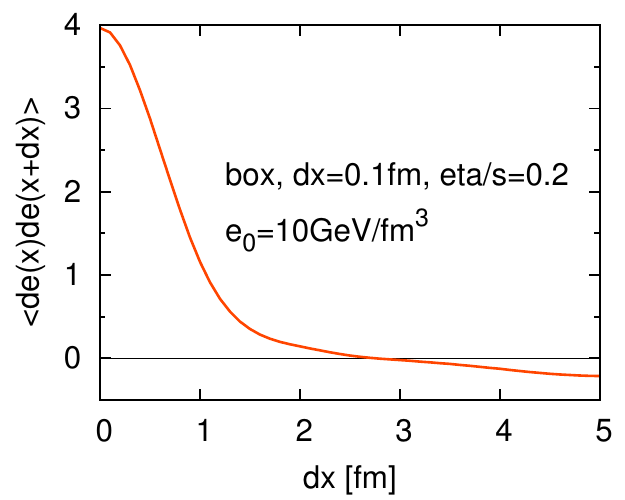}
\caption{Time evolution of the variance $\langle (\Delta e)^2\rangle$ (left) and correlation function 
$\langle \Delta e(x)\Delta e(x + dx)\rangle$ (right).}
\label{fig:variance}
\end{figure}

We initialize the energy density homogenously in the box at a value of $e_0=10$~GeV/fm$^3$. Now, the lattice spacing is given by $\Delta x = 0.1$~fm, which is an order of magnitude smaller and therefore strongly increases the variance of the noise field according to the delta-function in (\ref{eq:covariance}). This can cause large gradients, which the present algorithm cannot always handle. We, therefore, smooth the noise field over a correlation length of $1$~fm. In the left plot of Fig. \ref{fig:variance} we show how the variance of the energy density builds up after starting our calculations. Around $10$~fm/c one can see that the variance has saturated and fluctuates around its average value. We use averaging over timestep in order to analyse further quantities like the spatial correlation function of the energy density correlator, shown in the right plot of Fig. \ref{fig:variance}. For $dx=0$ this corresponds to the local variance. It can be observed that the correlation length of the noise 
field of $1$~fm is well reflected in the correlations of energy density. At larger distances one finds negative correlations due to total energy conservation.

Naively, one would expect that due to (\ref{eq:noiseav}), the thermal averages of thermodynamic quantities like the energy density should be the same, irrespectively of the presence of $\Xi^{\mu\nu}$ in the fluid dynamical equations (\ref{eq:fluideq}). It has been shown, however, that the nonlinearities present in the full fluid dynamical equations lead to nontrivial corrections \cite{Chafin:2012eq,Kovtun:2011np}. In the retarded shear-shear correlator, for example, one can identify three contributions stemming from the interactions of the fluctuations:
\begin{equation}
 G^{xyxy}_{R,{\rm shear-shear}}(\omega,\mathbf{0})=-\frac{7T}{90\pi^2}\Lambda^3-i\omega\frac{7T}{60\pi^2}\frac{\Lambda}{\gamma_\eta}+{(i+1)\omega^{3/2}\frac{7T}{90\pi^2}\frac{1}{\gamma_\eta^{3/2}}}\, ,
\label{eq:contr}
\end{equation}
where the first term on the RHS is a cutoff-dependent contribution to the equilibrium pressure, the second term is a cutoff-dependent contribution to the shear viscosity $\eta$ and the third term is a frequency-dependent contribution to both the shear viscosity $\eta$ and the relaxation time $\tau_\pi$.

\begin{figure}
 \includegraphics[width=0.49\textwidth]{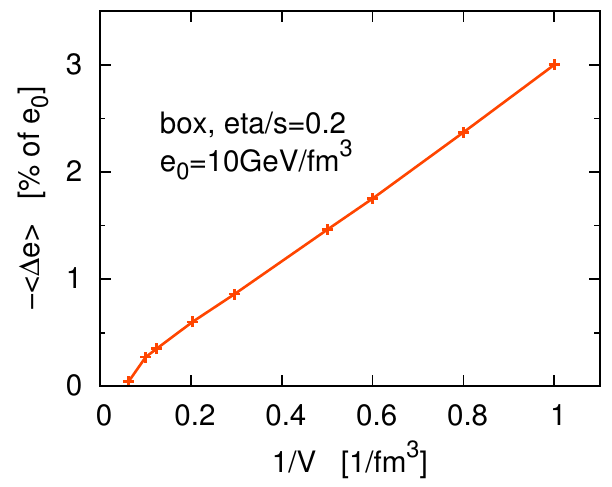}
 \includegraphics[width=0.49\textwidth]{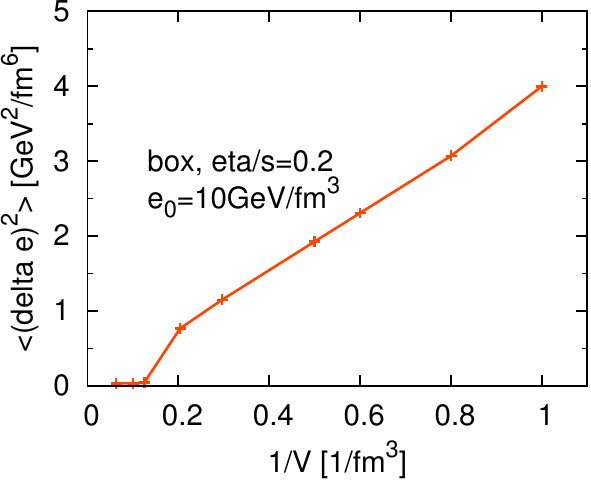}
\caption{Dependence on the lattice spacing $V^{1/3}$ of the correction to the average energy density due to interactions of fluctuations (left) and of the variance of energy density fluctuations (right).}
\label{fig:Vdep}
\end{figure}

In Fig. \ref{fig:Vdep} we vary the lattice spacing $\Delta x$ and therefore the cutoff parameter $\Lambda$ for values $>1$~fm, in order to avoid additional smoothing of the noise field and have the algorithm run stable. We observe that both the correction to the average energy density (left plot) as well as the variance of the energy density (right plot) scale as the third power (inversely proportional to the cell volume) of the cutoff parameter. We note, however, that the overall value of both quantities is only about $30-40$~\% of what is expected from (\ref{eq:contr}) and grandcanonical thermodynamics. This might be due to discretization effects in the numerical simulations and total energy conservation.

\section{Summary}
In the beginning era of precision measurements performed in heavy-ion collisions it becomes necessary to include refinements in the theoretical models that might lead to effects on the percent scale. One of these refinements is the fluid dynamical treatment of thermal fluctuations, which are of special interest at the QCD phase transition, in particular near the critical point, in small systems and for measurements of higher-order flow harmonics. We have presented recent progress on including these fluctuations in $3+1$d relativistic viscous fluid dynamics. It turns out that the interaction of fluctuations due to the nonlinearities inherent in the fluid dynamical equations leads to a renormalization of the equation of state and the transport coefficients. A systematic effort is needed to ensure the accurate propagation of fluctuations in fluid dynamical models.

\section*{Acknowledgments}
The authors thank Y.~Karpenko for providing the vHLLE code and useful discussions.
M.N. acknowledges support from a fellowship within the Postdoc-Program of the 
German Academic Exchange Service (DAAD). The work of M.B. is funded by the European 
Union's Horizon~2020 research and innovation programme under the Marie Sk\l{}odowska 
Curie grant agreement No 665778 via the National Science Center, Poland, under grant 
Polonez UMO-2016/21/P/ST2/04035. This work was supported in parts by the U.S. 
Department of Energy under grants DE-FG02-03ER41260 and DE-FG02-05ER41367. The authors 
acknowledge fruitful discussions within the Beam Energy Scan Theory (BEST) Topical 
Collaboration.


\begin{thebibliography}{99}

\bibitem{Nahrgang:2011mg}
  M.~Nahrgang, S.~Leupold, C.~Herold and M.~Bleicher,
  Phys.\ Rev.\ C {\bf 84} (2011) 024912.
  
\bibitem{Nahrgang:2011mv}
  M.~Nahrgang, S.~Leupold and M.~Bleicher,
  Phys.\ Lett.\ B {\bf 711} (2012) 109.
\bibitem{Nahrgang:2011vn}
  M.~Nahrgang, C.~Herold, S.~Leupold, I.~Mishustin and M.~Bleicher,
  J.\ Phys.\ G {\bf 40} (2013) 055108.
\bibitem{Herold:2013bi}
  C.~Herold, M.~Nahrgang, I.~Mishustin and M.~Bleicher,
  Phys.\ Rev.\ C {\bf 87} (2013) no.1,  014907.
\bibitem{Herold:2016uvv}
  C.~Herold, M.~Nahrgang, Y.~Yan and C.~Kobdaj,
  Phys.\ Rev.\ C {\bf 93} (2016) no.2,  021902.

\bibitem{Kapusta:2011gt}
  J.~I.~Kapusta, B.~M\"uller and M.~Stephanov,
  Phys.\ Rev.\ C {\bf 85} (2012) 054906.

\bibitem{Kapusta:2012zb}
  J.~I.~Kapusta and J.~M.~Torres-Rincon,
  Phys.\ Rev.\ C {\bf 86} (2012) 054911.

\bibitem{Akamatsu:2016llw}
  Y.~Akamatsu, A.~Mazeliauskas and D.~Teaney,
  Phys.\ Rev.\ C {\bf 95} (2017) no.1,  014909.
\bibitem{Young:2014pka}
  C.~Young, J.~I.~Kapusta, C.~Gale, S.~Jeon and B.~Schenke,
  Phys.\ Rev.\ C {\bf 91} (2015) no.4,  044901.

\bibitem{Murase:2016rhl}
  K.~Murase and T.~Hirano,
  Nucl.\ Phys.\ A {\bf 956} (2016) 276.


\bibitem{Karpenko:2013wva}
  I.~Karpenko, P.~Huovinen and M.~Bleicher,
  Comput.\ Phys.\ Commun.\  {\bf 185} (2014) 3016.
\bibitem{Chafin:2012eq}
  C.~Chafin and T.~Sch\"afer,
  Phys.\ Rev.\ A {\bf 87} (2013) no.2,  023629.

\bibitem{Kovtun:2011np}
  P.~Kovtun, G.~D.~Moore and P.~Romatschke,
  Phys.\ Rev.\ D {\bf 84} (2011) 025006.
\end{thebibliography}
\end{document}